\documentclass[alpha-refs]{wiley-article}
\usepackage{color}
\usepackage{ulem}

\newcommand{\delete}[1]{}

\newcommand{\mnote}[1]{}

\usepackage{siunitx}
\usepackage{subcaption}

\graphicspath{{./}}	% Root directory of the pictures: remove once submitted 

\papertype{Original Article}
\title{A strategy to compute convective timescales of the Indian monsoon with the WRF model}

\author[1,2]{Lucy G. Recchia}
\author[1,2]{Valerio Lucarini}
\affil[1]{Department of Mathematics and Statistics, University of Reading, Reading, UK}
\affil[2]{Centre for the Mathematics of Planet Earth, University of Reading, Reading, UK}

\corraddress{L. G. Recchia, Department of Mathematics and Statistics, University of Reading, Reading, RG6 6AX, UK}
\corremail{l.g.recchia@reading.ac.uk}

\fundinginfo{L. G. Recchia, Horizon 2020 EU project TiPES, Grant/Award Number: 820970 \& UK NERC Leeds-York DTP, Grant/Award Number: NE/L002574/1; V. Lucarini, Horizon 2020 EU project TiPES, Grant/Award Number: 820970, Marie Curie ITN CriticalEarth, Grant/Award Number: 956170 \& EPSRC, Grant/Award Number: EP/T018178/1.}

\runningauthor{Recchia \& Lucarini.}

\begin{document}

\maketitle

\begin{abstract}
The Indian monsoon brings around 80\% of the annual rainfall over the summer months June--September to the Indian subcontinent. The timing of the monsoon onset and the associated rainfall has a large impact on agriculture, thus impacting the livelihoods of over one billion people. To improve forecasting the monsoon on sub-seasonal timescales, global climate models are in continual development. One of the key issues is the representation of convection, which is typically parametrised. Different convection schemes offer varying degrees of performance, depending on the model and scenario. Here, we propose a method to compute a convective timescale, which could be used as a metric for comparison across different models and convection schemes. The method involves the determination of a vertical convective flux between the lower and upper troposphere through moisture budget analysis, and then relating this to the total column moisture content. The method is applied to a WRF model simulation of the 2016 Indian monsoon, giving convective timescales that are reduced by a factor of 2 when the onset of the monsoon occurs. The convective timescale can also be used as an indicator of monsoon transitions from pre-onset to full phase of the monsoon, and to assess changes in monsoon phases under future climate scenarios. 

% Please include a maximum of seven keywords
\end{abstract}

%-----------------------------------------------------------------------------------------------
%-----------------------------------------------------------------------------------------------
\section{Introduction}
\label{intro}

%------------------------------------------------
%\textbf{Background/motivation}\\
Every summer (June--September), a seasonal reversal of the wind pattern heralds the beginning of the Indian monsoon, bringing significant rainfall to the Indian subcontinent. Accurate prediction of the timing, location and intensity of rainfall resulting from the monsoon is important for agriculture, which over one billion people are dependent on for subsistence \citep{gadgil06}. Similarly, forewarning of early or late onsets can help inform crop planting schedules \citep{bombardi20}. Whilst representation of the physical processes associated with the development of Indian monsoon is improving in weather and climate models \citep{gusain20,choudhury22,khadka22}, accurate forecasting on a seasonal time scale remains difficult \citep{johnson17,mohanty19,chevuturi21}. 

%------------------------------------------------
%\textbf{Overview of monsoon dynamics}\\ 
The pre-monsoon period, running from March to May, is characterised by increasing temperatures and intermittent rainfall, linked with the development of low pressure systems over the Bay of Bengal which can evolve into intense, localised thunderstorms. The enhancement of the land-sea thermal contrast and the migration of the inter-tropical convergence zone to the north contribute to the initiation of the Indian monsoon. A reversal in the large-scale circulation causes an influx of moisture to the Indian subcontinent, via the low level southwesterly wind travelling across the Arabian Sea. This creates favourable conditions for shallow convection over the southern peninsula, which helps to moisten the lower troposphere and leads to the formation of shallow cumulus clouds \citep{parker16,menon18}. Onset is first declared in Kerala, typically around 1st June, once sufficient rainfall has fallen in the region, then the onset progresses to the northwest against the mid level wind field over the following 6 weeks. The speed of the propagation of the onset to the northwest is modulated by the surface soil moisture and intensity of the mid-tropospheric dry intrusion \citep{krishnamurti10,krishnamurti12,parker16,volonte20,menon22}. 

%------------------------------------------------
%\textbf{Monsoon phases: active spells \& breaks}\\ 
Once the monsoon onset has progressed over all of the subcontinent, the full monsoon season lasts until September, then withdraws. The full monsoon phase is characterised by active and break periods, based on anomalous rainfall over several consecutive days. The frequency of such periods typically determines a good or a poor monsoon year, and has significant impacts on the agricultural industry. Break periods are associated with a reduction in rainfall, weakened Somali jet, strengthened northwesterly winds over North-West \& Central India and weak convective activity \citep{krishnan00,gadgil03,rajeevan10,hannachi13}. On the other hand, increased rainfall, a stronger Somali jet, weakened northwesterly winds over North-West \& Central India, intensified convective activity and more frequent synoptic disturbances are the main features of active spells \citep{krishnan00,rajeevan10,hannachi13}. Predictability of active and break periods in the Indian monsoon is important because they often lead to severe weather events such as flooding or droughts. 

%------------------------------------------------
%\textbf{Convective regimes}\\
Different convective regimes, such as active and break periods of the Indian monsoon, can be categorised by a timescale for convective adjustment. Longer convective timescales (days) are representative of short-lived convective events, or ``triggered'' convection \citep{emanuel94,done06,zimmer11,molini11,keil11,keil14,bechtold14}, which are caused by local-scale disturbances. Shorter convective timescales (hours) are indicative of a statistical equilibrium regime, controlled by large-scale processes on relatively slow timescales and the rate of Convective Available Potential Energy (CAPE) creation to removal \citep{emanuel94,done06,zimmer11,molini11,keil11,keil14,bechtold14}. In terms of the Indian monsoon, longer convective timescales would be expected in the pre-monsoon period, where localised intermittent deep convection is prevalent and the Convective Inhibition (CIN) is high \citep{parker16,volonte20}. During the full monsoon, where the large-scale flow dominates, there is more widespread deep convection, high CAPE and low CIN, implying shorter convective timescales. 

%------------------------------------------------
%\textbf{Convective parametrisation}\\ 
The timescale difference between large-scale processes (slow) and convective response (fast) is a general assumption of most closures for bulk mass-flux convective parametrisation schemes \citep{bk:stensrud}, typically used in GCMs. Following \citet{mapes97bk}, closures can be grouped into deep-layer or low-level control schemes. The creation (and removal) of CAPE by large-scale processes is the focus of deep-layer control schemes, whilst low-level control schemes consider the reduction of CIN through boundary layer processes. Convective parametrisation schemes can also be categorised by their sensitivity primarily to moisture, or primarily to instability \citep{bk:stensrud}. For the moisture-control schemes, a key concept is determining the moisture availability from large-scale convergence and boundary layer turbulence, for example, and relating this to the convective activity. For convection to occur, certain conditions based on the column integrated moisture convergence or the amount of CIN must be met. 

A particular difficulty relates to the representation of convection in Global Climate Models (GCMs). Typically, the process of convection is parametrised, leading to wet or dry biases in the precipitation pattern \citep{mukhopadhyay10,willetts17}. There are multiple ways of parametrising convection, with each parametrisation scheme having its own advantages and disadvantages. The calculation of a convective timescale could be a useful tool both in marking the transition of the Indian monsoon from pre-onset to full monsoon, and as a metric for comparison between different convective parametrisation schemes and GCMs. 

%------------------------------------------------
%\textbf{Aerosols/climate change}\\
The value of provided a robust definition of a convective timescale for pre-onset and full monsoon periods could further be used to help assess changes to the Indian monsoon system under future climate scenarios. Precipitation associated with the monsoon is expected to increase, with more frequent extreme rainfall events \citep{moon20,ipcc6,monerie22}, highlighting the importance of accurate representation of the location and intensity of convection in GCMs. Additionally, break periods in the monsoon, linked with heatwaves and droughts, are likely to occur more often \citep{ipcc6}. The calculated convective timescale could be used to indicate break and active periods in GCM future climate simulations, as well helping to quantify any changes to climatological onset dates and monsoon season duration. A key aspect of uncertainty in future climate scenarios is the spread, concentration and composition of aerosols. This is particularly important for the Indian monsoon, as aerosols are known to increase the static stability of the atmosphere, suppressing moist convection and thus precipitation \citep{li16,wilcox20,ayantika21,recchia23}. Moreover, \citet{bollasina13c} shows that aerosols are responsible for an earlier monsoon onset in historical simulations. Further work is needed to help quantify future projections of the Indian monsoon as well as to unify the concepts behind and development of convective parametrisation schemes.  

%------------------------------------------------
%\textbf{Our strategy}\\
Here, a convective timescale is defined based on the relation between total column moisture and vertical convective flux. These quantities are integrated over a considerably sized region, in order to capture the large-scale processes responsible for determining the moisture budget of the atmosphere. Such a coarse-grained (yet adaptable and flexible) metric can be used for models of different level of complexity. For example, it could be used to compare different convective parametrisation schemes for a particular event, such as the Indian monsoon, as well indicating the type of convective regime. It can help quantify conditions based on the amount of convective activity, such as pre-onset and post-onset, and active/break periods during the full monsoon.

Section \ref{method} outlines a general method for calculating convective timescales using a moisture budget analysis. Section \ref{data} presents a case study of the Indian monsoon onset for the year 2016, as simulated by the Weather Research and Forecasting (WRF) model \citep{shamrock19}, which includes an assessment of the model performance. Section \ref{tconvwrf} shows the specific application of the definition of a convective timescale to the WRF model simulation of the 2016 Indian monsoon onset. In Section \ref{conc} we present a discussion of our results and suggest further lines of research.

%-----------------------------------------------------------------------------------------------
\section{Method for computing convective time scales}
\label{method}
We outline a strategy below for calculating convective time scales in weather and climate model simulations. Although the application here is restricted to the Indian monsoon, the method could be extended to other regions. The convective timescale is derived from the relationship between total column moisture and a vertical convective flux, where the vertical convective flux is the residual term in a moisture budget analysis over two layers of the atmosphere. The atmosphere is separated at the 700 hPa level, so that the cloud processes relating to rainfall are contained within the upper layer, following the assumptions for the idealised model presented in \citet{recchia21}.  

%------------------------------------------------
\subsection{Derivation of the vertical convective flux}
\label{method_convflux}
\begin{figure}[bt]
\centering
\includegraphics[width=12cm]{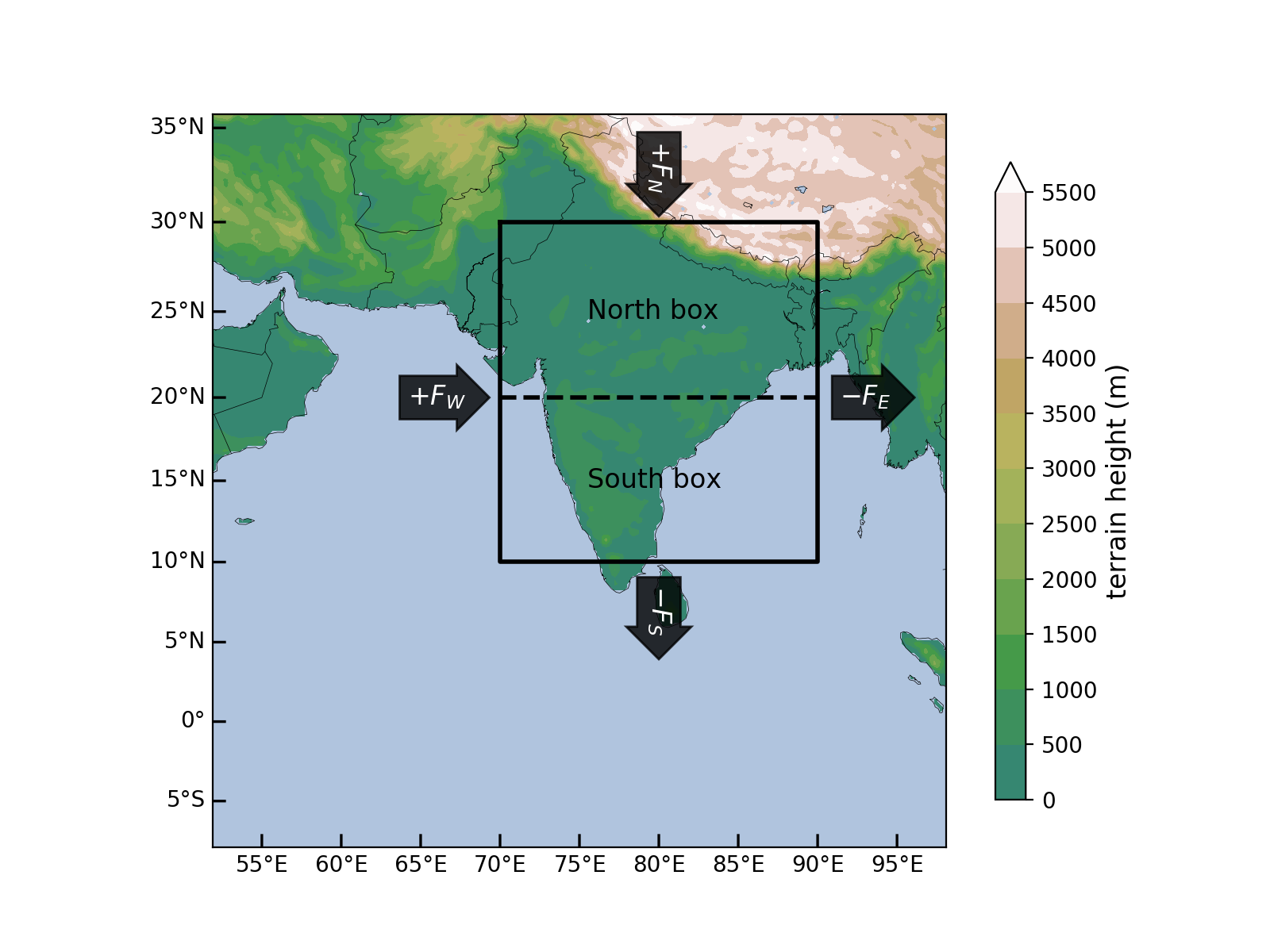} 
\caption{Regions over which the moisture budget is calculated (black box \& black dashed line), according to Equations \ref{equation01a} and \ref{equation01b}, including direction of fluxes at north, east, south and west sides. Figure area represents the simulation domain and shading shows terrain height (m).}
\label{figure01}
\end{figure}

We follow \citet{recchia20} and consider the moisture budget of two atmospheric layers. The upper layer is above 700 hPa and the lower layer extends from the surface to 700 hPa. For the case here, the moisture budget is computed for a region extending from 10--30$^\circ$N and 70--90$^\circ$E, encapsulating India. The horizontal bounds are denoted by the black box in Figure \ref{figure01}. Given the significant difference in behaviour between the northern and southern portions of the considered black box, the moisture budget and subsequent analysis is computed separately for each of these regions, with the division shown by the black dashed line in Figure \ref{figure01}. The equations below:

\begin{subequations}
\begin{align}
\textrm{Upper layer:}\ -F_{\textrm{conv}} &= -\Delta PW_{L2} + F_{N_{L2}} - F_{S_{L2}} + F_{W_{L2}} - F_{E_{L2}} - P. \label{equation01a} \\
\textrm{Lower layer:}\ +F_{\textrm{conv}} &= -\Delta PW_{L1} + F_{N_{L1}} - F_{S_{L1}} + F_{W_{L1}} - F_{E_{L1}} + E - q_{\textrm{sfc}}\omega_{\textrm{sfc}}. \label{equation01b}
\end{align}
\end{subequations}

describe the moisture budget, where $F_{\textrm{conv}}$ is the vertical convective moisture flux.  

Horizontal moisture fluxes ($F_N$, $F_E$, $F_S$, $F_W$) into the region are positive, whilst horizontal fluxes out of the region are negative, as shown by the white arrows in Figure \ref{figure01}. Precipitation ($P$) is contained solely within the upper layer and has a negative sign as it represents a loss of moisture. Evaporation ($E$) is a moisture source for the lower layer, and therefore has a positive sign. The integral of the vertical moisture flux convergence reduces to a surface term ($q_{\textrm{sfc}}\omega_{\textrm{sfc}}$); a moisture sink. The change in precipitable water (total column moisture) is signified by $\Delta PW$. All quantities are given in the standard mass flux units of kgm\textsuperscript{-2}s\textsuperscript{-1}.

Note that in some cases, numerical errors can lead to imperfect conservation of water across the layers \citep{lucarini07dan, lucarini08dan}. Thus, the average of the upper and lower layer vertical convective fluxes is taken to determine the overall vertical convective flux.

Equations \ref{equation01a} and \ref{equation01b} can be derived from first principles, considering the change in moisture over time for a fixed volume, applying mass balance and mass conservation of water vapour and resolving vertical and horizontal integrals. See also \citet{bk:peixoto} for an accurate treatment of water budgets in different spatial and temporal domains.

%------------------------------------------------
\subsection{Calculating the convective time scale}
\label{method_convtime}
A convective timescale can be computed by relating the variability of the total column moisture in both the upper and lower layers ($PW_{L2}$ and $PW_{L1}$) to the variability of the vertical convective flux ($F_{\textrm{conv}}$). We perform a linear fit between the anomaly ($\Delta$) in the total column moisture and the anomaly in the vertical convective flux. We have: 

\begin{equation}
T_{\textrm{conv}} = \frac{\Delta\left(PW_{L1} + PW_{L2}\right)}{\Delta F_{\textrm{conv}}}. \label{equation02} 
\end{equation}

Observations support the importance of total column moisture particularly in the lower troposphere, for the development of tropical deep convection (e.g. \citet{holloway09}). In practice, a scatter plot is used examine the correlation between the total column moisture in the layers and the vertical convective flux. The units of the total column moisture and the convective flux are kgm\textsuperscript{-2} and kgm\textsuperscript{-2}s\textsuperscript{-1} respectively, giving units of seconds (s) for the convective timescale. This time scale quantifies the efficiency in the process of moving water across the layers of the atmosphere.

%-----------------------------------------------------------------------------------------------
\section{WRF model simulation}
\label{data}
The WRF model \citep{shamrock19} is used to simulate the Indian summer monsoon for the year 2016. WRF is one of the most widely used numerical weather prediction models, and is employed internationally for both research and operational purposes \citep{powers17}. It is freely available, relatively easy to use and computationally efficient. 

The set up of the WRF model for our simulation is given below, including the combination of parametrisation schemes used. Then, the simulation is described and evaluated against reanalysis and observational data, focusing on the low and mid-level moisture development and timing of the transitions from pre-onset through to full monsoon.  

%------------------------------------------------
\subsection{WRF model set-up}
\label{data_model}
The 11 week simulation with the WRF model runs from 15th May 2016 to 1st August 2016, capturing the onset of the Indian monsoon and the transition from pre-onset to full monsoon. The reference year, 2016, is chosen for its concurrence with the INteraction of Convective Organisation with Monsoon Precipitation, Atmosphere, Surface and Sea (INCOMPASS) field campaign \citep{turner20}. 

%----------------
\begin{table}[bt]
\caption{Combination of physics scheme options for the tropical suite in ARW model.}
\label{table01} 
\begin{tabular}{lll} \headrow
\thead{Physics scheme type} & \thead{Name of scheme} & \thead{Reference} \\ 
Longwave radiation        & RRTMG                            & \citet{iacono08} \\
Shortwave radiation       & RRTMG shortwave                  & \citet{iacono08} \\
Microphysics              & WSM6                             & \citet{hong06a} \\
Cumulus parametrisation  & New Tiedtke                      & \citet{zhang11} \\ 
Planetary boundary layer  & Yonsei University                & \citet{hong06b} \\
Land-atmosphere interface & Noah Land Surface Model          & \citet{tewari04} \\
Surface layer             & MM5 similarity (old)             & \citet{monin54}; \\
                          &                                  & \citet{paulson70, webb70}; \\
                          &                                  & \citet{dyer70}; \\
                          &                                  & \citet{zhang82}; \\
                          &                                  & \citet{beljaars94} \\ \hline
\end{tabular}
\end{table}
%----------------
%%
%%
%------------------------------------------------
%\textbf{Technical details}\\
The WRF model version is the Advanced Research WRF 4.0 \citep{shamrock19}, used with the tropical suite, which sets the physics schemes (see Table \ref{table01}). A 60s timestep is chosen with data being output every 6 hours. A regional configuration is employed, with the domain bounds being 52.5--97.5$^\circ$E, 7.5$^\circ$S--37.5$^\circ$N, on a $20\times 20$ km horizontal grid. The lateral boundaries are forced at 6-hourly intervals using the NCEP GDAS/FNL atmospheric dataset ds083.3 \citep{ncep_data}. Terrestrial data is derived from MODIS satellite products \citep{modis_data, friedl02}. The sea surface temperature, derived from the skin temperature in the atmospheric dataset, is kept constant throughout the simulation. 

%------------------------------------------------
\subsection{Description of the 2016 Indian monsoon simulation}
\label{data_descrip}
\begin{figure}[bt]
\centering
\includegraphics[width=12cm]{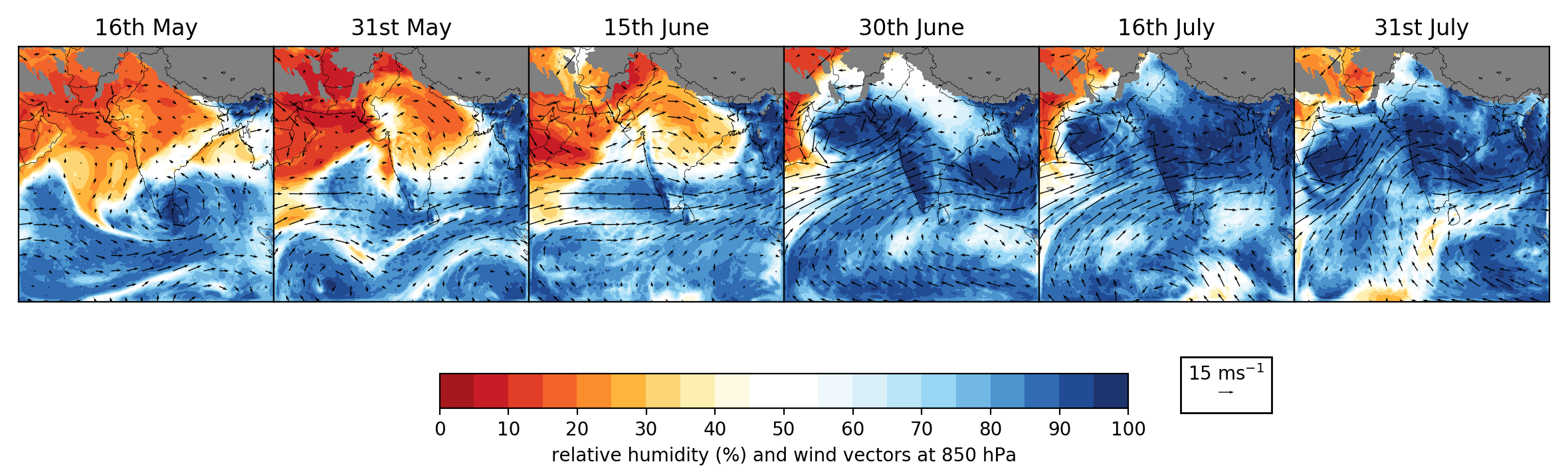} 
\caption{Daily averaged relative humidity (shading, \%) and wind (vectors, ms\textsuperscript{-1}), at the 850 hPa level, for various dates from the 11 week simulation with the WRF model. Areas of high orography are masked in grey.}
\label{figure02}
\end{figure}

\begin{figure}[bt]
\centering
\includegraphics[width=12cm]{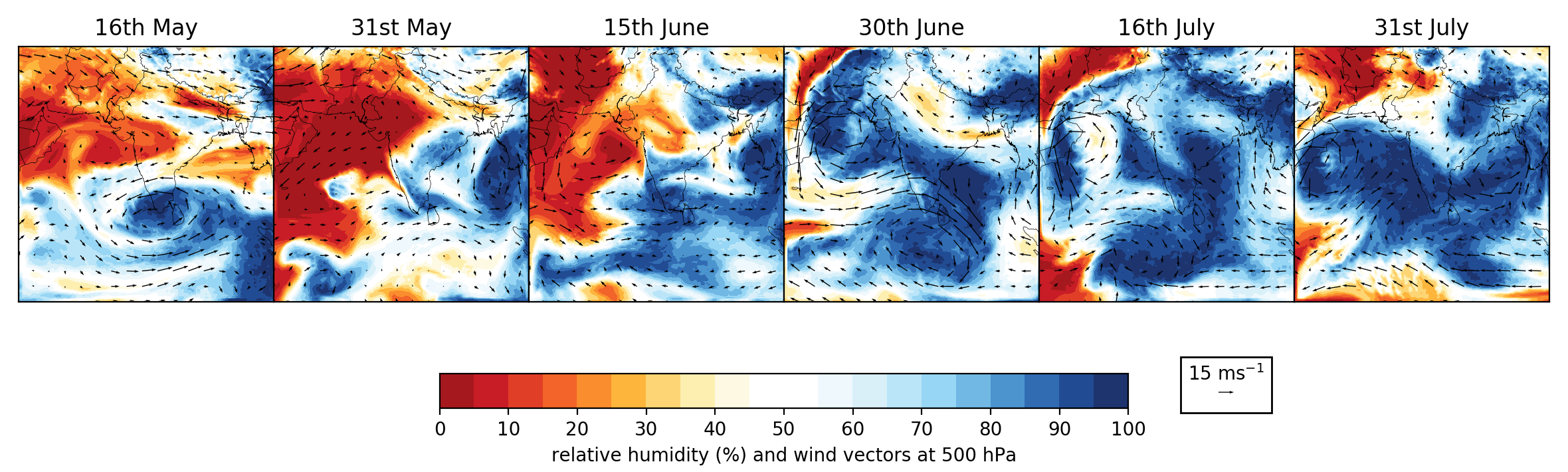} 
\caption{Daily averaged relative humidity (shading, \%) and wind (vectors, ms\textsuperscript{-1}), at the 500 hPa level, for various dates from the 11 week simulation with the WRF model. Areas of high orography are masked in grey.}
\label{figure03}
\end{figure}

\begin{figure}[bt]
\centering
\includegraphics[width=12cm]{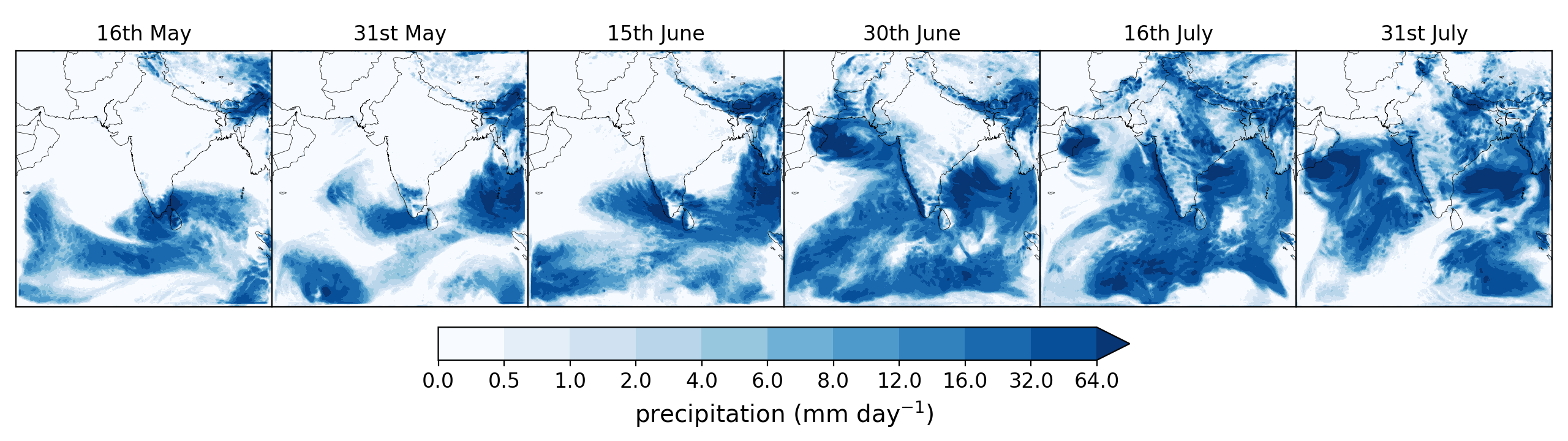}
\caption{Daily accumulated precipitation (mm/day), for various dates from the 11 week simulation with the WRF model.}
\label{figure04}
\end{figure}

The 11 week simulation can be categorised by phases of the Indian monsoon: pre-onset, first onset and transition to full monsoon. The first onset occurs in the southwestern state of Kerala, and is typically declared on or around the 1st June. Over the following 6 weeks, the onset progresses to the northwest so that by the end of July, the monsoon covers the entire Indian subcontinent. During the 2016 monsoon season, a late onset was observed (8th June), as well as monsoon depressions over the Bay of Bengal and Arabian Sea, respectively. Figures \ref{figure02}--\ref{figure05} and Figures \ref{figureS01}--\ref{figureS04} show the synoptic conditions during the phases of the Indian monsoon for the 11 week simulation. 

%--------------------------
\subsubsection{Pre-onset}
\label{data_descrip_pre}
The first two weeks of the WRF model simulation, corresponding to the last two weeks in May 2016, exhibit the typical meteorological conditions prior to the onset of the monsoon. Across the Indian landmass, surface temperatures range from 25--30$^\circ$ in the south, to over 35$^\circ$ in the centre and north (Figure \ref{figureS01}). The surrounding ocean is approximately 5$^\circ$ cooler than the majority of the land, setting up a land-sea thermal contrast. In conjunction with the temperature gradient, a pressure gradient also develops as the Intertropical Convergence Zone migrates northwards, forming a region of low pressure that extends horizontally from north India to the Arabian Sea along the Himalayan foothills, and vertically from low to mid-levels. Figure \ref{figureS02} shows the evolution of this low pressure region, the monsoon trough, as it strengthens throughout May and into June.

The majority of India experiences arid conditions, with little rainfall and low relative humidity - see Figures \ref{figure02}--\ref{figure04}. Only the southern peninsula and oceans receive rainfall. Throughout May, warm, dry air is driven from the Eurasian continent to India. Towards the end of May, the low-level southwesterly monsoon flow (The Somali Jet) begins to strengthen and deepen, bringing moisture to south India. The boundary between continental dry and oceanic moist air masses is clearly visible at mid-levels in the 31st May panel (Figure \ref{figure03}). 

%--------------------------
\subsubsection{First onset}
\label{data_descrip_onset}
\begin{figure}[bt] 
\centering
\includegraphics[width=12cm]{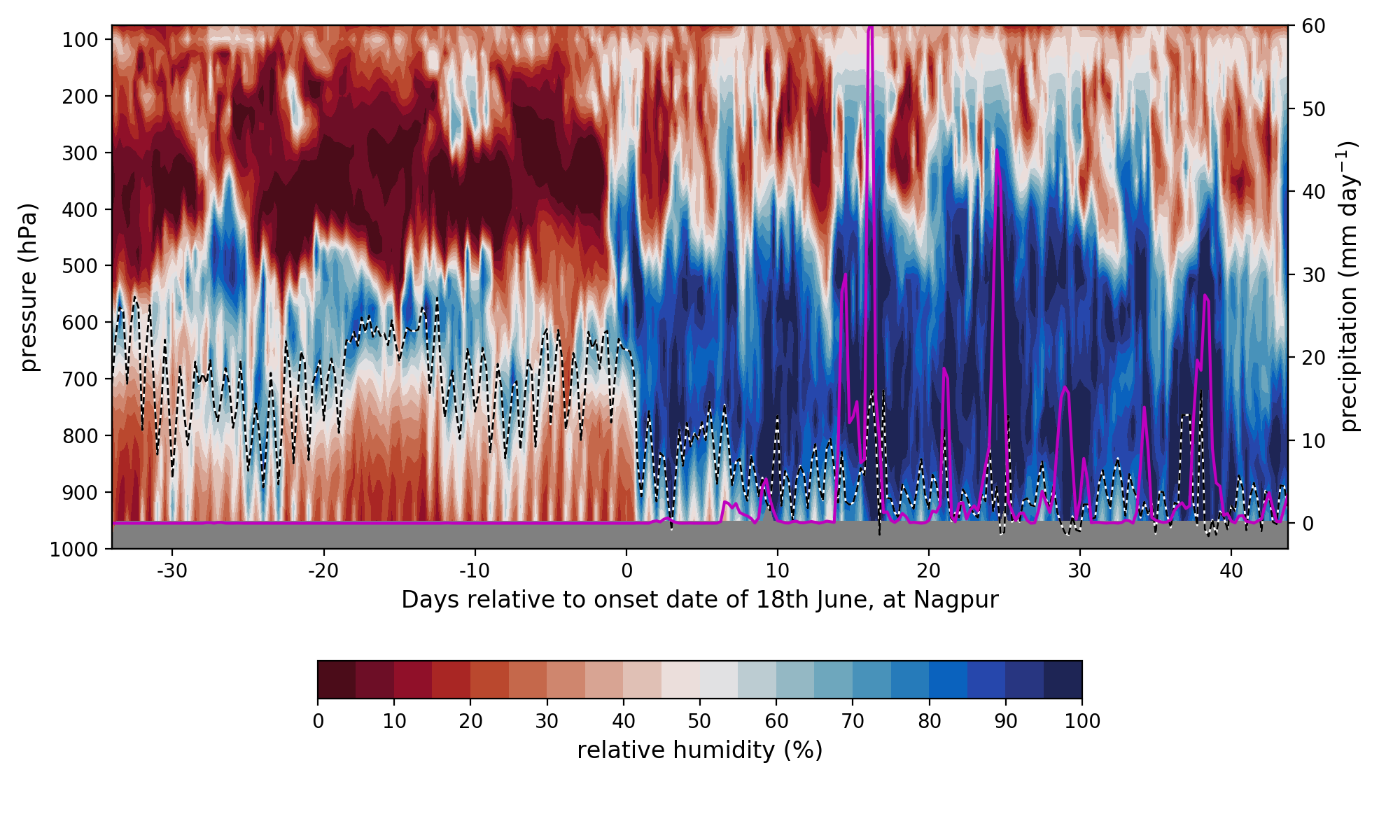} 
\caption{Time-pressure section of relative humidity for Nagpur, Central India, from 11 week simulation with the WRF model. Lifting condensation level (dashed line) and precipitation (magenta line) over-plotted. Onset date for Nagpur in 2016 is taken as 18th June. Areas of high orography are masked in grey.}
\label{figure05}
\end{figure}

The relative humidity over India remains low at the start of June (Figure \ref{figure02}), with moist air and rainfall confined to the southern peninsula, indicating a late onset. In this simulation, onset dates are estimated from Figures \ref{figure05} and \ref{figureS04}, which show the evolution of the relative humidity in an atmospheric column corresponding to the cities of Nagpur, Maharashtra and Kochi, Kerala, respectively. The first onset date of the Indian monsoon is estimated as 6th June at Kochi, Kerala in the simulation (Figure \ref{figureS04}), compared to the actual date of 8th June for the year 2016. After onset, the mid-level relative humidity ($\sim$500-700 hPa) increases by around 20\% (Figures \ref{figure03}, \ref{figureS04}). According to climatology, as mentioned above, the first onset is declared in Kerala in early June and by mid-June, the onset would have progressed to central India. In this simulation, the progression of the onset from southeast to northwest India is delayed, with little advancement between 31st May and 15th June, reflecting the real-world conditions during the 2016 monsoon season. 

Although the conditions at the end of May and mid-June are reasonably similar in terms of relative humidity and the large-scale wind field, there are some key differences. By 15th June, the low-level southwesterly monsoonal flow is more developed, having increased in speed and depth over the previous two weeks (Figure \ref{figure02}). The monsoon flow brings moisture from the Arabian Sea to the east coast and continues across India. At mid-levels, the dry air mass intruding from northwest to central India has begun to dissipate, in conjunction with a weakening of the northwesterly winds in the region (Figure \ref{figure03}). During mid-June at high levels (not shown), the westerly Subtropical Jet weakens and moves northwards; by the end of June, the Subtropical Jet has moved out of the simulation domain. In contrast, over the first two weeks of June, the Tropical Easterly Jet passing over central India has formed and strengthened. Considering Figure \ref{figureS03}, the gradual accumulation of soil moisture in south India implies a slight increase in local rainfall in the first two weeks of June, compared to May.

%--------------------------
\subsubsection{Transition to full monsoon}
\label{data_descrip_full}
The greatest change in the atmospheric state occurs between the middle and end of June, when the monsoon onset progresses very rapidly to the northwest of the Indian subcontinent. The simulated 2016 monsoon takes approximately 3 weeks to advance from first onset in Kerala to a state of full monsoon, compared to a usual progression time of six weeks. The dramatic increase in moisture at the time of onset for Nagpur in central India, estimated as the 18th June, is shown clearly by Figure \ref{figure05}. In addition, the lifting condensation level (dashed line) lowers after onset. Figure \ref{figure05} highlights the strong diurnal cycle at this location. An increase in rainfall at Nagpur (magenta line) can also be seen approximately a week after onset; in the month prior to the onset of the monsoon, there was no rainfall in Nagpur. The onset for Kerala and the southern peninsula is less sharp than for central and northwestern India, which has a relatively drier climate. 

The wedge of moist air associated with the monsoon extends from the surface to around 400 hPa. This sudden change in the relative humidity from $\sim$20\% to $\sim$80\% can also be seen in Figure \ref{figure03}, between the 15th June and the 30th June panels. A mid-level dry intrusion from the northwest is seen to persist at the end of June, but by mid-July, the opposing moist monsoon flow dominates. At low-levels, the southwesterly monsoon flow reaches its peak speed and continues to provide an influx of moisture from the Arabian Sea to India. 

The increased levels of atmospheric moisture after onset corresponds to increased precipitation, particularly over areas of high orography such as the Western Ghats on the west coast and the northeastern states of India (Figure \ref{figure04}). At the end of June, a large region of north India and Pakistan remains dry with little rainfall, due to the presence of a dry intrusion. By July, widespread precipitation is evident across the subcontinent. The progression of the monsoon onset and the effect of the dry intrusion is reflected in the soil moisture in Figure \ref{figureS03}. Between 15th and 30th June, a greater extent of soil is moistened, from the southern tip to central India, with a correlating decrease in surface temperature (Figure \ref{figureS01}). As the dry intrusion dissipates, soil moisture across central and northeast India increases, with the land becoming progressively wetter and cooler throughout July as the monsoon rains fall. 

Towards the end of June, a strong low pressure system (or monsoon depression) forms over the north of the Arabian Sea, with a weaker low pressure system also forming in the Bay of Bengal. The dark colours in Figure \ref{figureS02} and areas of cyclonic circulation in Figure \ref{figure02} indicate these low pressure systems. The additional rainfall associated with these low pressure systems can be seen in Figure \ref{figure04}. By the end of July, the monsoon depression over the Arabian Sea has weakened, whilst the low pressure system in the Bay of Bengal persists and travels inland towards India, then eastwards. 

%------------------------------------------------
\subsection{Performance of the WRF model simulation} 
\label{data_perf}
\begin{figure}[bt]
\centering 
\includegraphics[width=10cm]{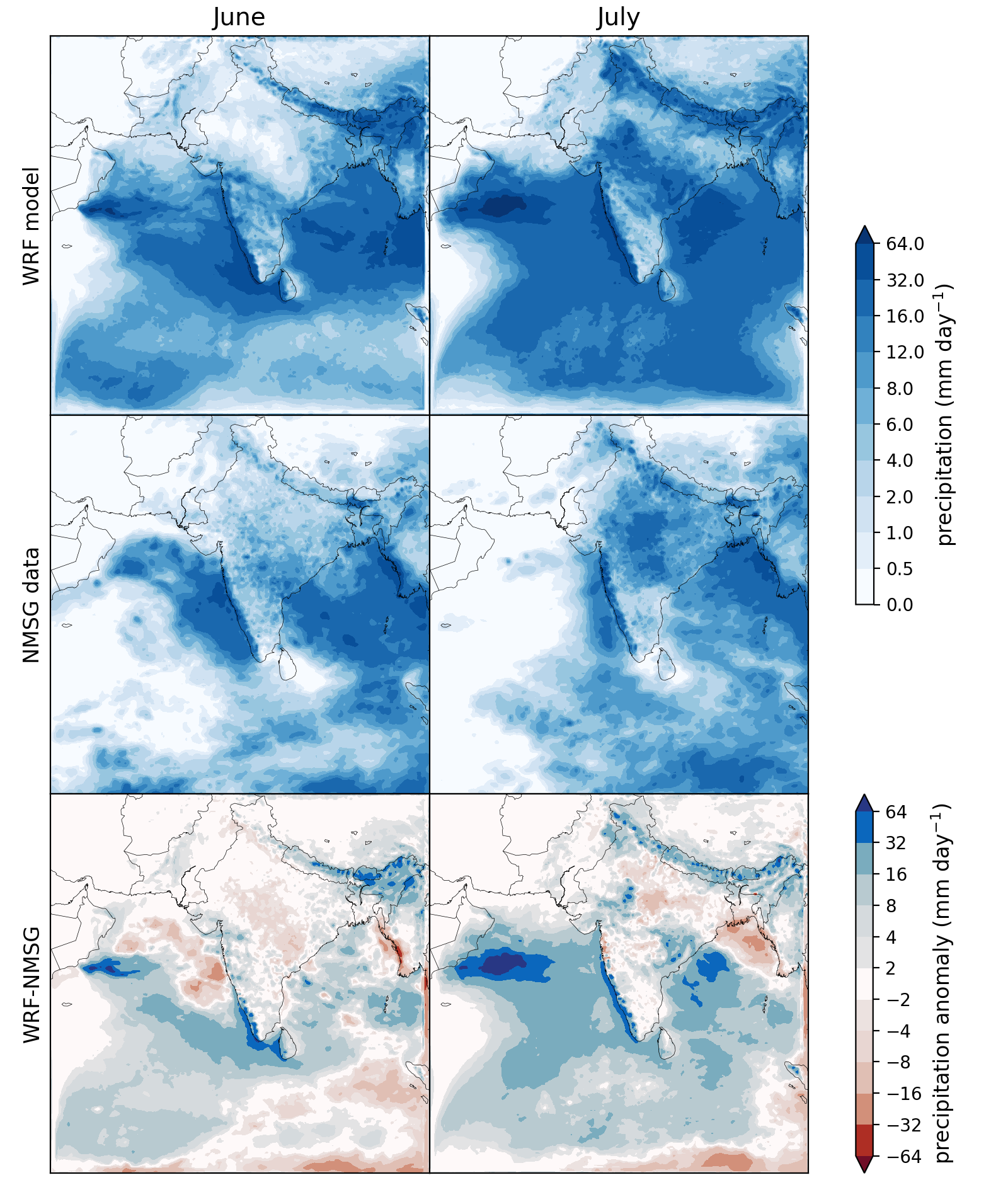} 
\caption{Monthly averaged daily accumulated precipitation (mm/day) for June \& July 2016. The top row shows data from the WRF model simulation and the middle row shows merged rain gauge/TRMM satellite data \citep{mitra09, mitra13}. The bottom row shows the anomaly between model and observed data.}
\label{figure06}
\end{figure}

A brief comparison between the WRF model simulation and various reanalysis and observational datasets is given, focusing on the representation of precipitation, relative humidity and circulation. A merged dataset (referred to as NMSG) derived from rain gauge and the TRMM satellite measurements, described in \citet{mitra09, mitra13}, is used as a benchmark for the output precipitation field from the WRF model. Figure \ref{figure06} shows the daily accumulated precipitation, averaged over June and July for the WRF model (top row), the merged NMSG data (middle row) and the anomaly (bottom row). In general, the WRF model produces too much rainfall over the oceans, particularly during July. Rainfall associated with the low pressure system over the Arabian Sea appears to be anomalous compared with the NMSG dataset. Over northwest India and into central India, the WRF model slightly underestimates the precipitation. In contrast, the precipitation is overestimated by around 32 mm/day offshore of the Western Ghats. The WRF model also produces more precipitation than is observed along the Himalayan mountain range, especially to the east. Overall, the WRF model does a reasonable job of reproducing the observed precipitation field over India, despite some localised regions of exaggerated intensity.

Next, the relative humidity and large-scale circulation are considered against two reanalysis datasets, ERA5 \citep{era5_data} and NCEP GDAS/FNL \citep{ncep_data}. Note that whilst the NCEP GDAS/FNL dataset is used to set the lateral boundaries, the interior of the domain evolves freely.

\begin{figure}[bt] 
\centering
\includegraphics[width=10cm]{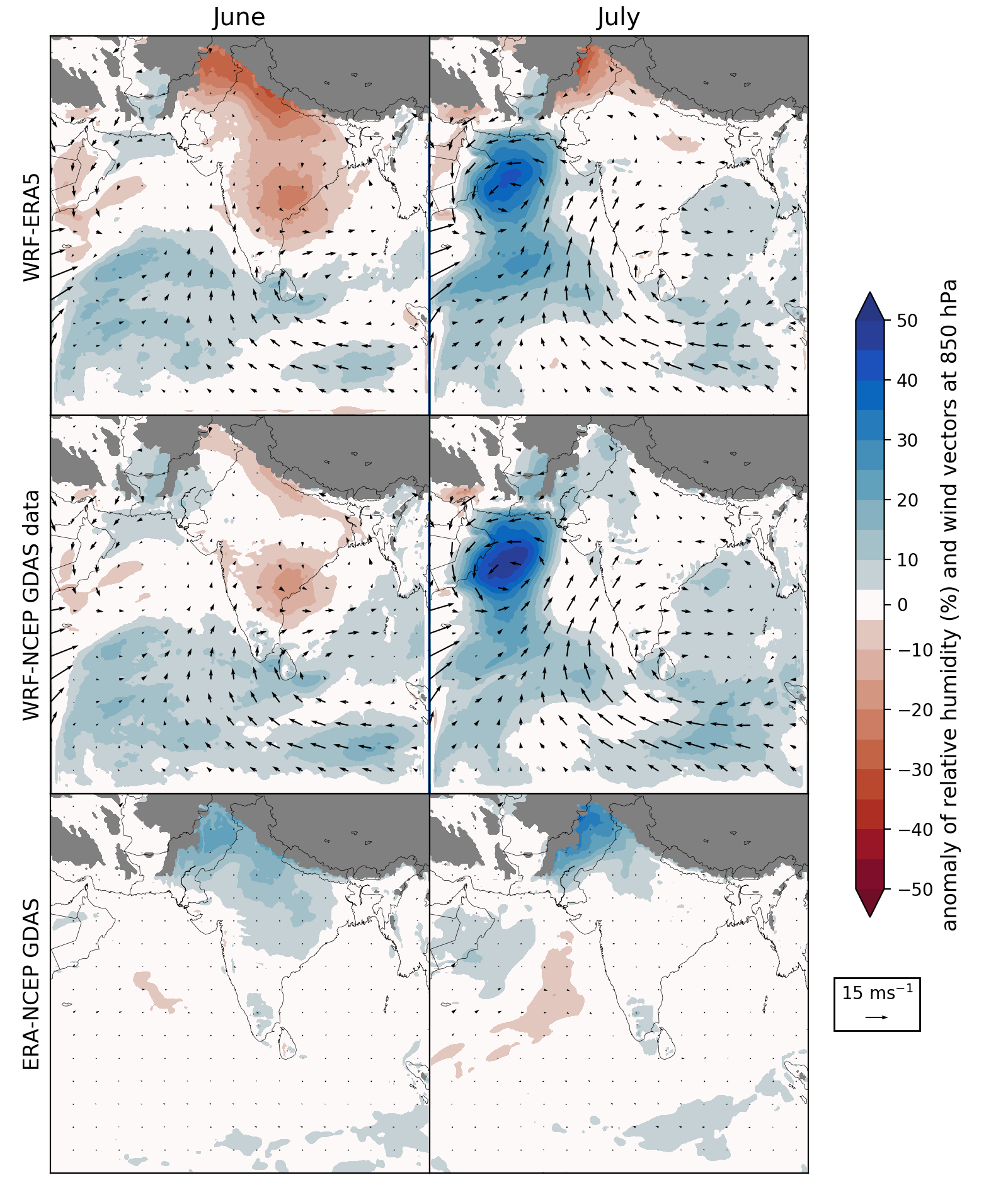} 
\caption{Anomaly of monthly averaged relative humidity (\%) and wind vectors (ms\textsuperscript{-1}) at 850 hPa for June \& July 2016 between the WRF model \& ERA5 reanalysis data (top row), the WRF model \& NCEP GDAS/FNL reanalysis data (middle row) and ERA5 \& NCEP GDAS/FNL reanalysis data (bottom row).}
\label{figure07}
\end{figure}

At low levels (Figure \ref{figure07}, top row), the WRF model shows a dry bias over northwest–central India for June, which is slightly more pronounced when compared against ERA5 then NCEP GDAS/FNL data. In July, the anomaly with NCEP GDAS/FNL data shows a wet bias over the north Arabian Sea (Figure \ref{figure07}, middle row), which is associated with a low pressure system. The northwesterly dry intrusion is a much more prominent feature during July in the NCEP GDAS/FNL dataset than in the ERA5 dataset (Figure \ref{figure07}, bottom row). 

The WRF model overestimates the relative humidity at 850 hPa over the oceans by 5--10\%, and also overestimates the low-level wind speed. These factors contribute to the overestimation of precipitation over the oceans. At 500 hPa (Figure \ref{figureS05}), the wet bias over the oceans in the WRF model simulation is enhanced. Compared to both reanalysis datasets, the WRF model is generally drier over Central India and more moist over South India. In June, the dry bias extends from Central India towards the Arabian Sea, whilst in July the dry bias extends in the oppositr direction, from Central India to the Bay of Bengal. The WRF model produces slightly stronger easterly winds at mid levels in June and July to the south of India, compared to either of the reanalysis datasets. 

The ERA5 and NCEP GDAS/FNL datasets agree reasonably well for both June and July, at mid-levels as well as low-levels, with the most relevant difference being the underestimation of the intensity of the northwesterly dry intrusion in ERA5. At high levels (200 hPa, not shown), the WRF simulation has a significant dry bias over the domain for June and July, compared to either of the reanalysis datasets.  

The monsoon depression that occurs over the Arabian Sea during late June--July in the WRF simulation (described in Section \ref{data_descrip_full}) is not evident in either of the reanalysis datasets. \citet{imdweb} note the occurrence of a monsoon depression over the Arabian Sea at the end of June 2016, although it is weaker and dissipates much faster in real-life than in the WRF simulation. Similarly, observations from the INCOMPASS field campaign confirm the presence of another monsoon depression, which develops over the Bay of Bengal at the start of July and propagates inland towards northwest India \citep{martin20}.

The WRF simulation of the 2016 Indian summer monsoon is in fair agreement with ERA5 and NCEP GDAS/FNL reanalysis data, and is qualitatively consistent with observations from the INCOMPASS field campaign and the corresponding Met Office Unified Model runs \citep{fletcher18,turner20,volonte20,martin20}. Most importantly, the WRF simulation reproduces the timing of the transitions between pre-onset, mid-onset and the full monsoon, despite variations in the moisture content and horizontal wind field over the interior of India. In particular, the late onset at Kerala and the moistening of the troposphere which enables rapid advancement during mid–late June, are well-captured. 

%-----------------------------------------------------------------------------------------------
\section{Computing convective timescales of the Indian monsoon with the WRF model}
\label{tconvwrf}
We now compute the convective timescales for the dataset produced by the WRF model simulation of the 2016 Indian monsoon. The convective timescales for the northern and southern regions of the Indian subcontinent are calculated for the time periods before and after onset. It is expected that the convective timescale will be shorter during the full monsoon than pre-onset, given the higher convective activity and corresponding inverse relationship between the vertical convective flux and the convective timescale. 

%------------------------------------------------
\subsection{Moisture budget analysis}
\label{tconvwrf_budget}
Following the method outlined in Section \ref{method_convflux}, the moisture fluxes at the boundaries of the north and south boxes in Figure \ref{figure01} are computed for each layer. The residual term in the moisture budget gives the convective flux, as per Equations \ref{equation01a} and \ref{equation01b}. 

\begin{figure}[bt]
\centering 
\includegraphics[width=12cm]{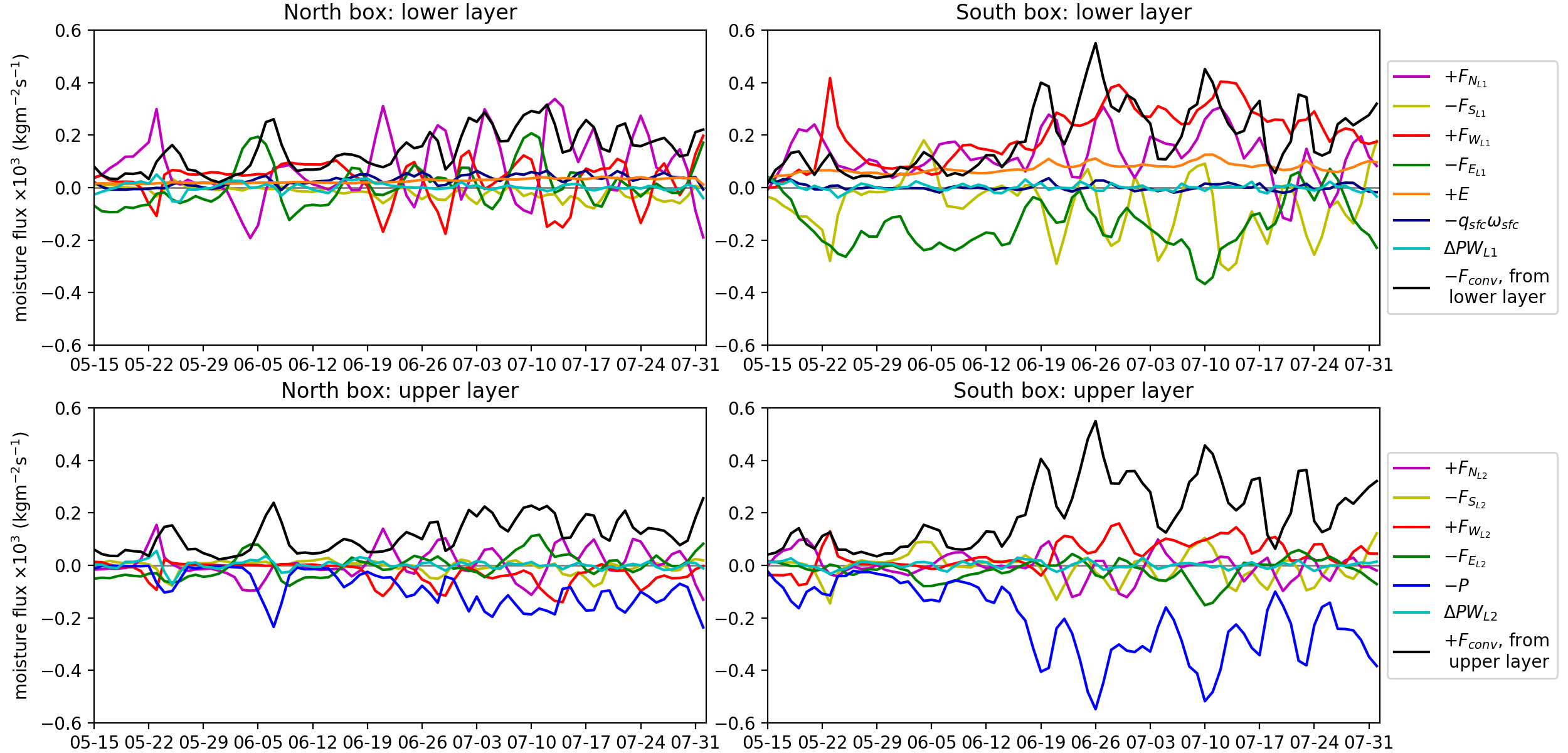}
\caption{Moisture budget components from 11 week simulation with the WRF model (daily averaged data), over the north \& south boxes shown in Figure \ref{figure01}, following Equations \ref{equation01a} and \ref{equation01b}. Dates ($x$-axis) in MM-DD format. Components have been interpolated to pressure levels and split into two layers at 700 hPa.}
\label{figure08}
\end{figure}

Figure \ref{figure08} shows the evolution of the upper and lower layer components over the 2016 monsoon season simulation, which has been produced using daily averaged data. It is evident that the southern region receives a greater amount of precipitation (blue line) than the northern region. The highest fluxes are also seen in the southern region; specifically, the lower layer western flux (red line), representing the moist inflow from the Arabian Sea, and the lower layer eastern flux (green line), representing the corresponding outflow towards southeast Asia. 

The convective flux (black line) is mostly determined by the precipitation, which is the largest component in the moisture budget. Similarly to the precipitation, the convective flux is greater over the southern region than the northern region. A significant increase in the convective flux and precipitation over the southern region can be seen from the 6th June, when the monsoon first onsets. A slight increase in the horizontal moisture fluxes also occurs around this time. For the northern region, the convective flux and precipitation become higher after onset, which has a later date of 18th June, but increase is less pronounced than for the southern region. The convective flux calculated from the lower layer moisture budget closely matches the convective flux calculated from the upper layer moisture budget, giving confidence in the method. Going forward, the average of the absolute value of the convective flux from the lower and upper layers is used. 

%------------------------------------------------
\subsection{Convective timescales for pre-onset and full monsoon}
\label{tconvwrf_monsoon}
\begin{figure}[bt]
\centering 
\includegraphics[width=12cm]{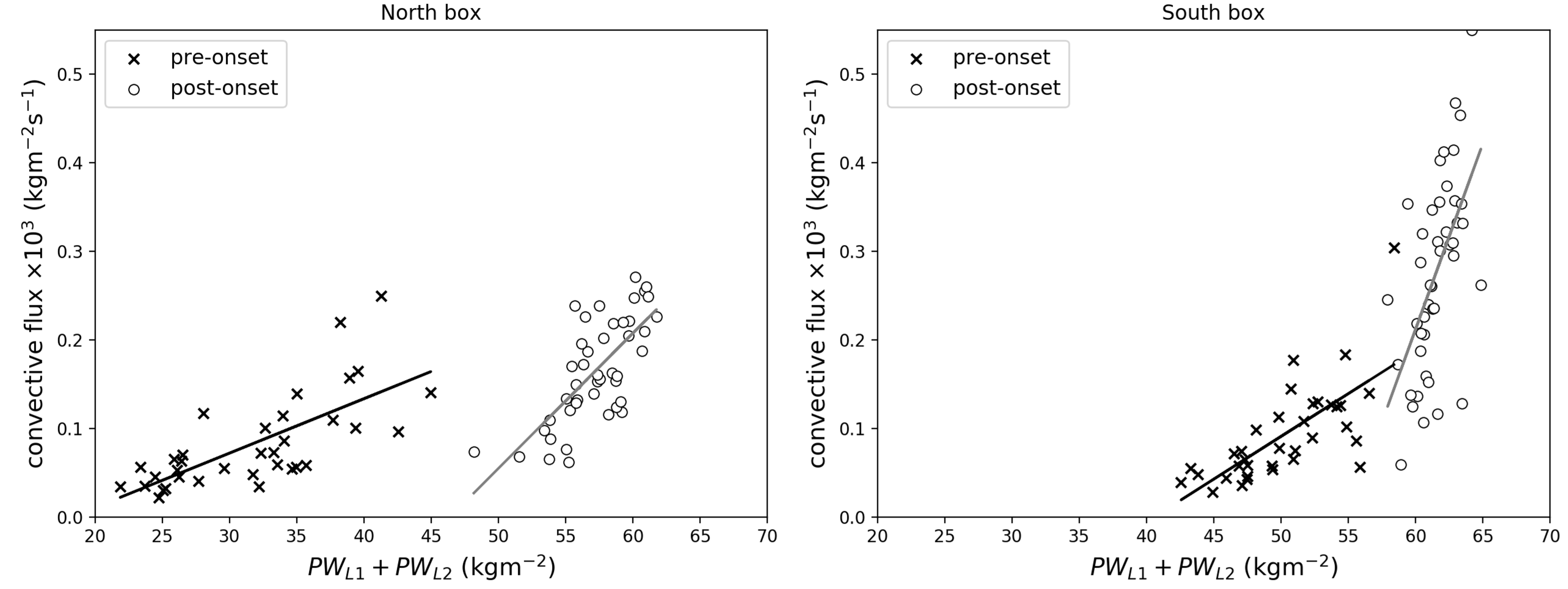} 
\caption{Correlation of vertical convective flux with the sum of the integrated moisture content in the lower and upper layers. Pre-onset refers to the period before 18th June, and post-onset the period after. Daily averaged data from 11 week simulation with the WRF model, over the north \& south boxes shown in Figure \ref{figure01}.}
\label{figure09}
\end{figure}

Figure \ref{figure09} clearly shows presence of a positive correlation between total column moisture ($PW_{L1}+PW_{L2}$) and convective flux ($F_{\textrm{conv}}$), for northern and southern regions, before and after the onset of the monsoon. The onset date of 18th June is taken for both regions.   

The onset of the monsoon is more dramatic in the northern region, with a clear transition in the conditions pre- and post-onset. In contrast, the onset over the southern is more gradual, with total column moisture and convective activity increasing in conjunction. This difference between northern and southern regions can also be seen in Figures \ref{figure05} and \ref{figureS04}, and is reflective of the higher levels of moisture and convective activity in the southern region during the pre-onset period, compared to the relatively dry and desert-like state of a large portion of the northern region.

After the onset of the monsoon, the total column moisture increases by 50\% or more and the convective flux increases slightly over the northern region, and greatly over the southern region. The increase in the slope of the regression line from pre-onset to post-onset indicates a reduction in the convective timescale, which corresponds to more and/or shorter convective events occurring after the monsoon has onset. 

If the difference in moisture content between upper and lower layers is considered instead of the total column moisture, we find no significant correlation between the difference in layer moisture and the convective flux \citep{recchia20}. This is in contrast to \citet{recchia21}, who found that a convective flux based on the difference in moisture content between upper and lower layers gave more realistic results in an idealised model framework.

The convective timescale is determined using Equation \ref{equation02}. Results are shown in Table \ref{table02}, along with statistical parameters used to quantify the correlation between convective flux and total column moisture as presented in Figure \ref{figure09}. The convective timescale after the monsoon has onset is less than half the value prior to onset, for both regions, indicating that convective efficiency has more than doubled. Both pre- and post-onset convective timescales are shorter for the southern region than the northern region, again reflecting the higher levels of convective activity observed over the southern region. 

\begin{table}[bt]
\caption{Convective timescales, correlation coefficients ($r$) and p-values from the scatter plots in Figure \ref{figure09}.}
\label{table02}
\begin{tabular}{lcccr} \headrow
\thead{Region} & \thead{Time period} & \thead{$\boldsymbol T_{\textrm{conv}}$ (days)} & \thead{$\boldsymbol r$} & \thead{p-value} \\ 
North India  & pre-onset  & 1.88 & 0.712 & $<$0.01 \\
North India  & post-onset & 0.76 & 0.722 & $<$0.01 \\ 
South India  & pre-onset  & 1.20 & 0.708 & $<$0.01 \\
South India  & post-onset & 0.28 & 0.586 & $<$0.01 \\ 
All India    & pre-onset  & 1.49 & 0.784 & $<$0.01 \\
All India    & post-onset & 0.82 & 0.383 &    0.01 \\ \hline         
\end{tabular} 
\end{table}

For the idealised model presented by \citet{recchia21}, a convective timescale was assumed to be in the range of 0.5--7 days. The derived convective timescales from the WRF model simulations suggest a smaller range of 0.5--2 days. The correlation between total column moisture and convective flux is confirmed as strongly positive in all cases by the correlation coefficient ($r$) being in the range 0.5--1. Similarly, the p-values below 0.01 for each case gives a confidence level of 99\% in the results, indicating that our proposal for defining the convective timescale is supported by data.

%------------------------------------------------
\subsection{Sensitivity of the convective timescale to shallow \& deep convection} 
\label{tconvwrf_sens}
The sensitivity of the convective timescale to shallow and deep convection is assessed by comparing values between model simulations where either the shallow or deep convection has been switched off. Although it is difficult to analyse the differences in the simulations with convection switched off because the redistribution of moisture is unclear, there are several observations that can be made (Figures \ref{figureS06}--\ref{figureS08}). With no deep convection parametrisation, the precipitation pattern changes to more intense bursts over smaller, more localised regions. The relative humidity at 850 hPa is slightly reduced over the ocean when shallow \& mid level convection is switched off. At 500 hPa, the atmosphere is considerably drier when deep convection is switched off, compared to the control simulation. 

Repeating the moisture budget analysis and following the method described in Section \ref{tconvwrf}, the calculated convective timescales are presented in Tables \ref{table03} and \ref{table04}. Generally, with either shallow or mid convection switched off, the convective timescales are longer, particularly for North India prior to onset, as a result of the reduced ability to transfer water vertically. The convective timescales are slightly longer when there is no deep convection, compared to when there is no shallow or mid level convection. Nonetheless, regardless of the choice of the parametrisation, the convective timescale is able to distinguish very clearly between the active and inactive phase of the monsoon.

\begin{table}[bt] 
\caption{Convective timescales, correlation coefficients ($r$) and p-values from the scatter plots with shallow \& mid level convection switched off (Figure \ref{figureS09}).}
\label{table03}
\begin{tabular}{lcccr} \headrow
\thead{Region} & \thead{Time period} & \thead{$\boldsymbol T_{\textrm{conv}}$ (days)} & \thead{$\boldsymbol r$} & \thead{p-value} \\ 
North India  & pre-onset  & 4.55 & 0.564 & $<$0.01 \\
North India  & post-onset & 0.76 & 0.744 & $<$0.01 \\
South India  & pre-onset  & 1.03 & 0.820 & $<$0.01 \\
South India  & post-onset & 0.42 & 0.586 & $<$0.01 \\ 
All India    & pre-onset  & 1.82 & 0.745 & $<$0.01 \\
All India    & post-onset & 0.92 & 0.482 & $<$0.01 \\ \hline         
\end{tabular} 
\end{table}

\begin{table}[bt] 
\caption{Convective timescales, correlation coefficients ($r$) and p-values from the scatter plots with deep convection switched off (Figure \ref{figureS10}).}
\label{table04}
\begin{tabular}{lcccr} \headrow
\thead{Region} & \thead{Time period} & \thead{$\boldsymbol T_{\textrm{conv}}$ (days)} & \thead{$\boldsymbol r$} & \thead{p-value} \\ 
North India  & pre-onset  & 5.27 & 0.571 & $<$0.01 \\
North India  & post-onset & 0.96 & 0.703 & $<$0.01 \\ 
South India  & pre-onset  & 1.34 & 0.596 & $<$0.01 \\
South India  & post-onset & 0.43 & 0.527 & $<$0.01 \\ 
All India    & pre-onset  & 3.37 & 0.448 & $<$0.01 \\
All India    & post-onset & 1.42 & 0.436 & $<$0.01 \\ \hline         
\end{tabular} 
\end{table}

%-----------------------------------------------------------------------------------------------
\section{Conclusions} 
\label{conc}
A method to compute convective timescales for a GCM has been presented, with specific application to a WRF model simulation of the 2016 Indian monsoon. The derived convective timescales are a useful indicator of monsoon onset, with the convective timescale approximately halving pre- and post-onset. The deviation of a convective timescale has implications for convective adjustment in parametrised convective schemes, and could be used a metric for comparison across different GCMS. Applying the method presented here systematically to other GCMs, such as CMIP6-standard models \citep{eyring16}, would provide a greater level of information regarding the value and variance of the convective timescale during the Indian monsoon.

The WRF model is chosen to simulate the 2016 Indian monsoon over an 11 week period from mid-May to August, focusing on the on the onset of the monsoon and the transition to full monsoon conditions. The late onset and subsequent rapid progression from southeast to northwest India of the 2016 season are well captured by the WRF simulation, despite biases in the precipitation, humidity and circulation compared to reanalysis datasets. The ability of the WRF model to accurately represent the timing of the 2016 monsoon phases from pre- to post-onset was the most important factor in continuing further analysis. 

A convective timescale can be derived from GCM simulations, firstly by computing the vertical convective flux, which is inversely related to the convective timescale. A moisture budget over the region of interest, India, is constructed, separating the atmosphere into lower (surface--700 hPa) and upper (700--50 hPa) layers. The residual term from the total column moisture content, horizontal fluxes, precipitation and evaporation is the vertical convective flux. The convective timescale can then be determined by the rate of change of total layer moisture content to vertical convective flux. Applying this method to the WRF simulation, a scatter plot is used to relate the total layer moisture content to vertical convective flux. The points are separated into two time periods: pre-onset and post-onset. There is a robust relationship between total column moisture and vertical convective flux, which are positively correlated for both monsoon phases. A regression line is used to determine the rate of change, and thus the convective flux, both prior to and after monsoon onset.

Prior to monsoon onset, the convective timescales calculated for regions of North India, South India and All India, are in the range of 1--2 days. The convective timescales approximately halve after the monsoon has onset. The timescales for North India are slightly longer than South India, indicating slightly less convective activity in the North, as would be expected. If either shallow and mid convection or deep convection is switched off, the convective timescales become longer, particularly for North India. Removing deep convection has a greater impact on the convective timescales than removing shallow and mid convection. The calculated convective timescales are consistent with assumed values in the two-layer model presented in \citet{recchia21}, linking theoretical understanding with a practical method to quantify convective timescales, which are difficult to observe in the real-world monsoon. 

The convective timescale is a useful metric for model comparison and assessment of monsoon regimes. Furthermore, it could aid quantification of changes in monsoon conditions and timings under future climate scenarios; our proposed timescale could be a valuable addition to the standard metrics of Earth System Model Evaluation Tools for regional climatic features \citep{weigel21}. For example, the convective timescale provides a measurable way to quantify the effect of absorbing aerosols, which are known to suppress convection, on the rate of convective activity associated with the monsoon. The method presented here is adaptable to different scales, from weather to climate applications. Further work could examine whether the convective timescale provides an early warning signal of extreme events, such as localised heavy precipitation or droughts, which are expected to become more frequent in the future. Given the generality of the method of calculated presented here, it can also be applied to other regions and phenomena, such as the West African monsoon.       

%-----------------------------------------------------------------------------------------------
%-----------------------------------------------------------------------------------------------
\section*{Acknowledgements}
This project is TiPES contribution \#263: This project has received funding from the European Union’s Horizon 2020 research and innovation programme under grant agreement No 820970. L. G Recchia acknowledges the National Environmental Research Council, for funding through its Doctoral Training Partnership (Leeds-York). V. Lucarini acknowledges the support provided by the Marie Curie ITN CriticalEarth project and the Engineering and Physical Sciences Research Council. We also acknowledge Copernicus Climate Change Service, National Centers for Environmental Prediction, National Center for Atmospheric Research and Ashis K. Mitra for providing access to the various datasets used in our analysis. Further thanks to Douglas Parker for conceptualisation. 
%\add{We thank the two anonymous reviewers for their constructive comments and suggestions to improve the paper.} 

%\section*{Conflict of interest}
%You may be asked to provide a conflict of interest statement during the submission process. Please check the journal's author guidelines for details on what to include in this section. Please ensure you liaise with all co-authors to confirm agreement with the final statement.

\printendnotes

%-----------------------------------------------------------------------------------------------
\clearpage
\newpage
\section*{Supplementary figures}
%Supporting information will be included with the published article. For submission any supporting information should be supplied as separate files but referred to in the text.

\setcounter{page}{1}
\setcounter{figure}{0}
\makeatletter 
\renewcommand{\thefigure}{S\@arabic\c@figure}
\makeatother

\begin{figure}[!ht]
\centering
\includegraphics[width=12cm]{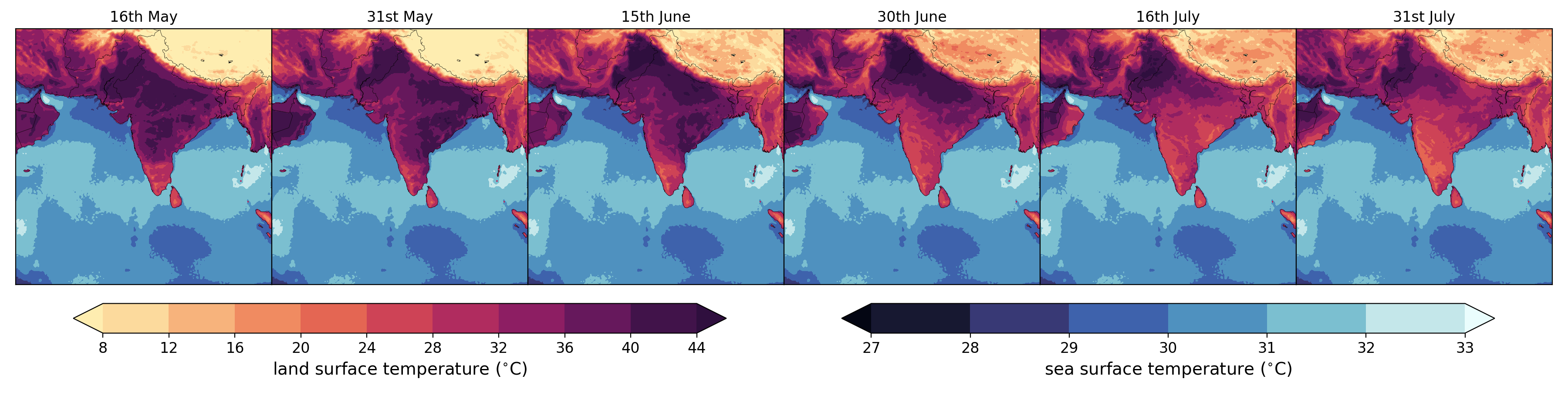}  
\caption{Daily averaged surface temperature ($^\circ$C), for various dates from the 11 week simulation with the WRF model. Areas of high orography are masked in grey.}
\label{figureS01}
\end{figure}

\begin{figure}[!ht]
\centering
\includegraphics[width=12cm]{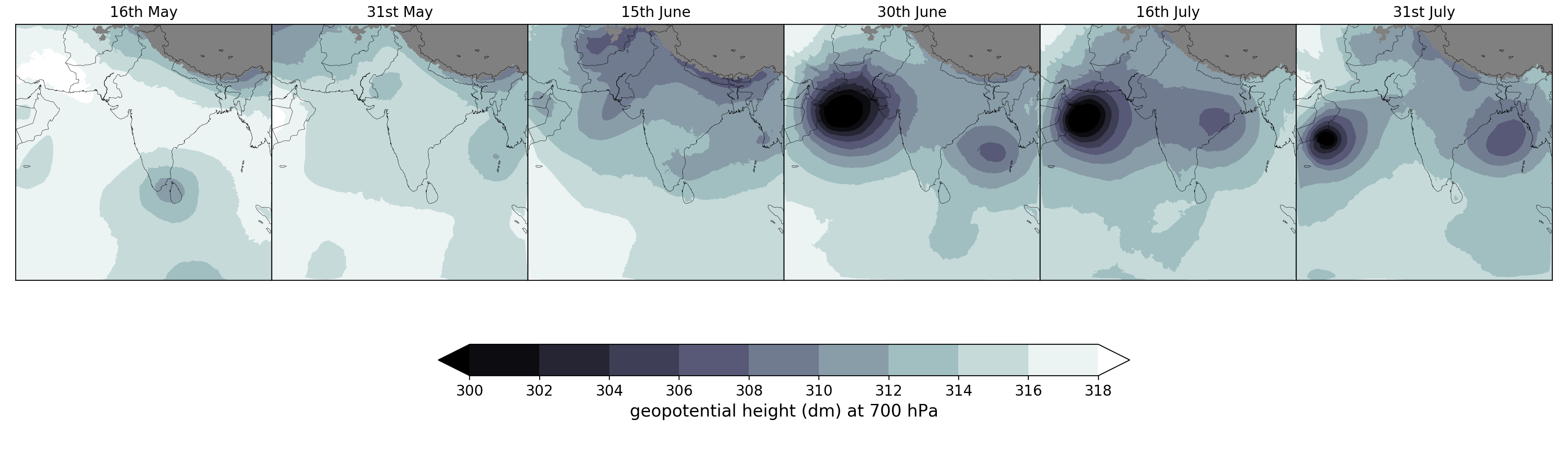} 
\caption{Geopotential height (dm) at the 700 hPa level, for various dates from the 11 week simulation with the WRF model. Areas of high orography are masked in grey.}
\label{figureS02}
\end{figure}

\begin{figure}[!ht]
\centering
\includegraphics[width=12cm]{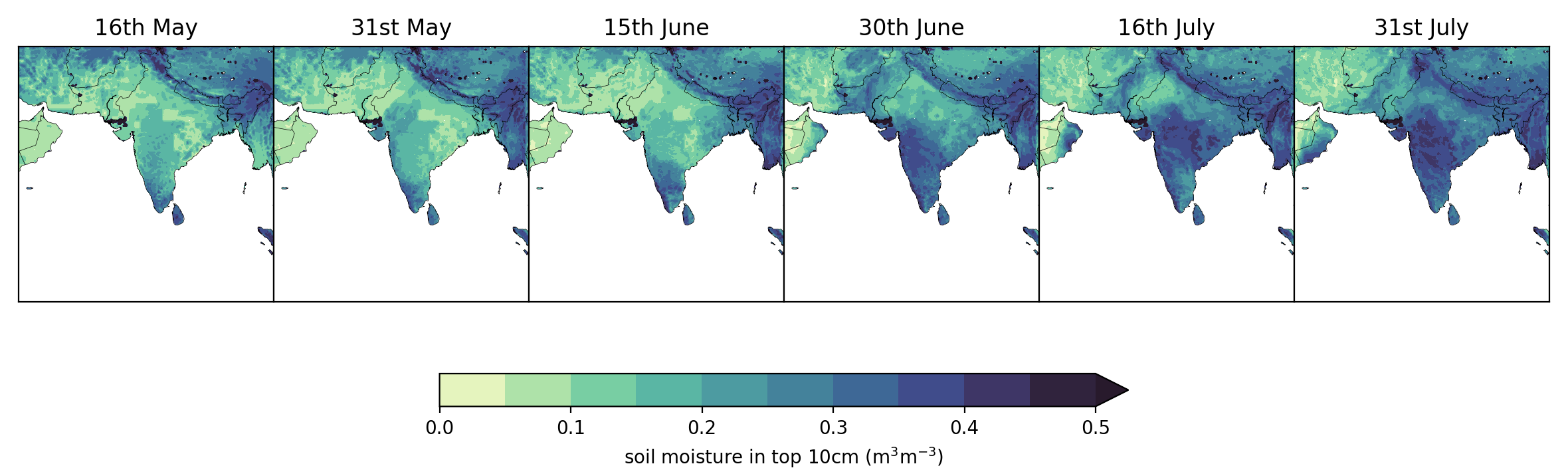} 
\caption{Daily averaged soil moisture (m\textsuperscript{3}m\textsuperscript{-3}) in the top 10cm, for various dates from the 11 week simulation with the WRF model.}
\label{figureS03}
\end{figure}

\begin{figure}[!ht]
\centering 
\includegraphics[width=10cm]{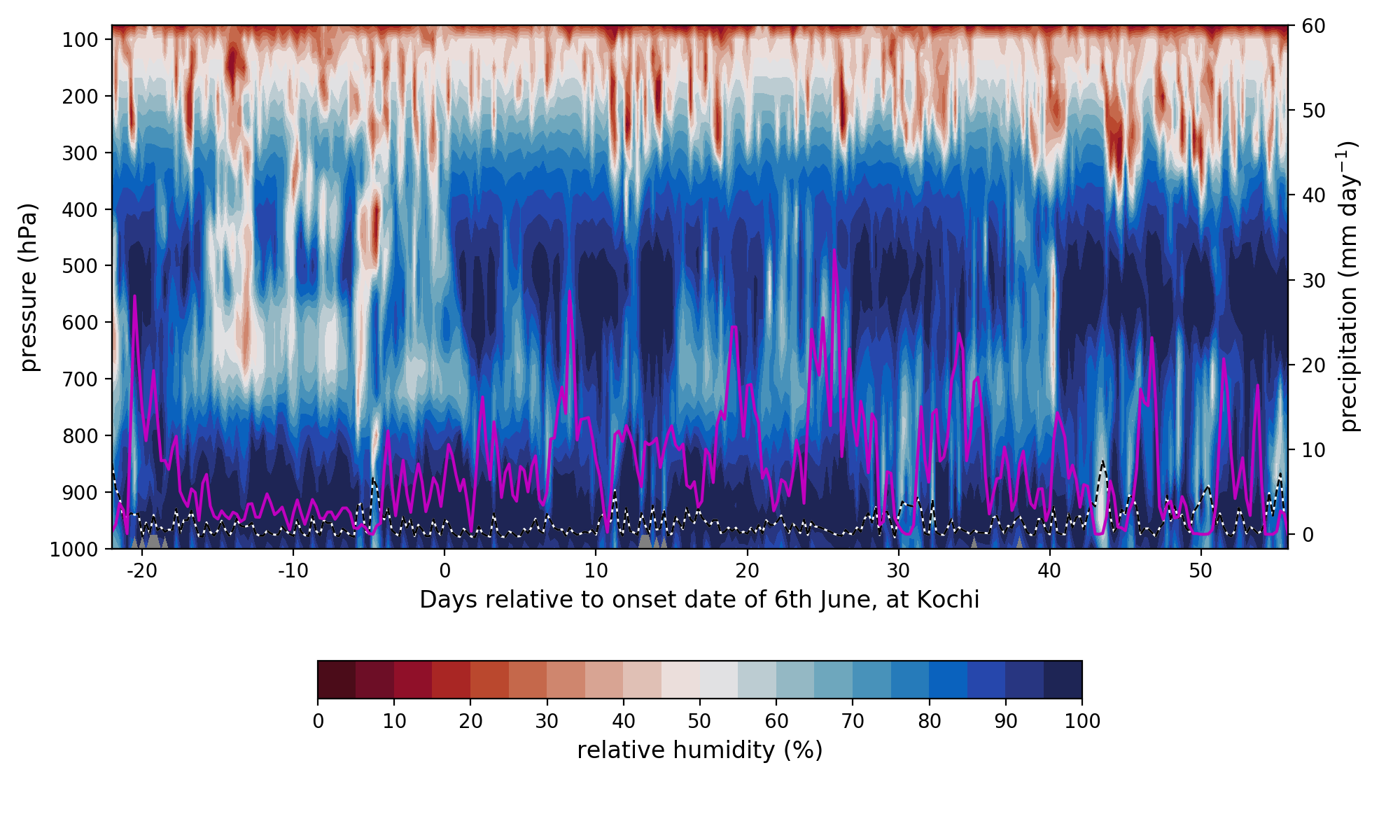} 
\caption{Time-pressure section of relative humidity for Kochi, Kerala, from 11 week simulation with the WRF model. Lifting condensation level (dashed line) and precipitation (magenta line) over-plotted. Onset date for Kochi in 2016 is taken as 6th June. Areas of high orography are masked in grey.}
\label{figureS04}
\end{figure}

\begin{figure}[!ht]
\centering 
\includegraphics[width=10cm]{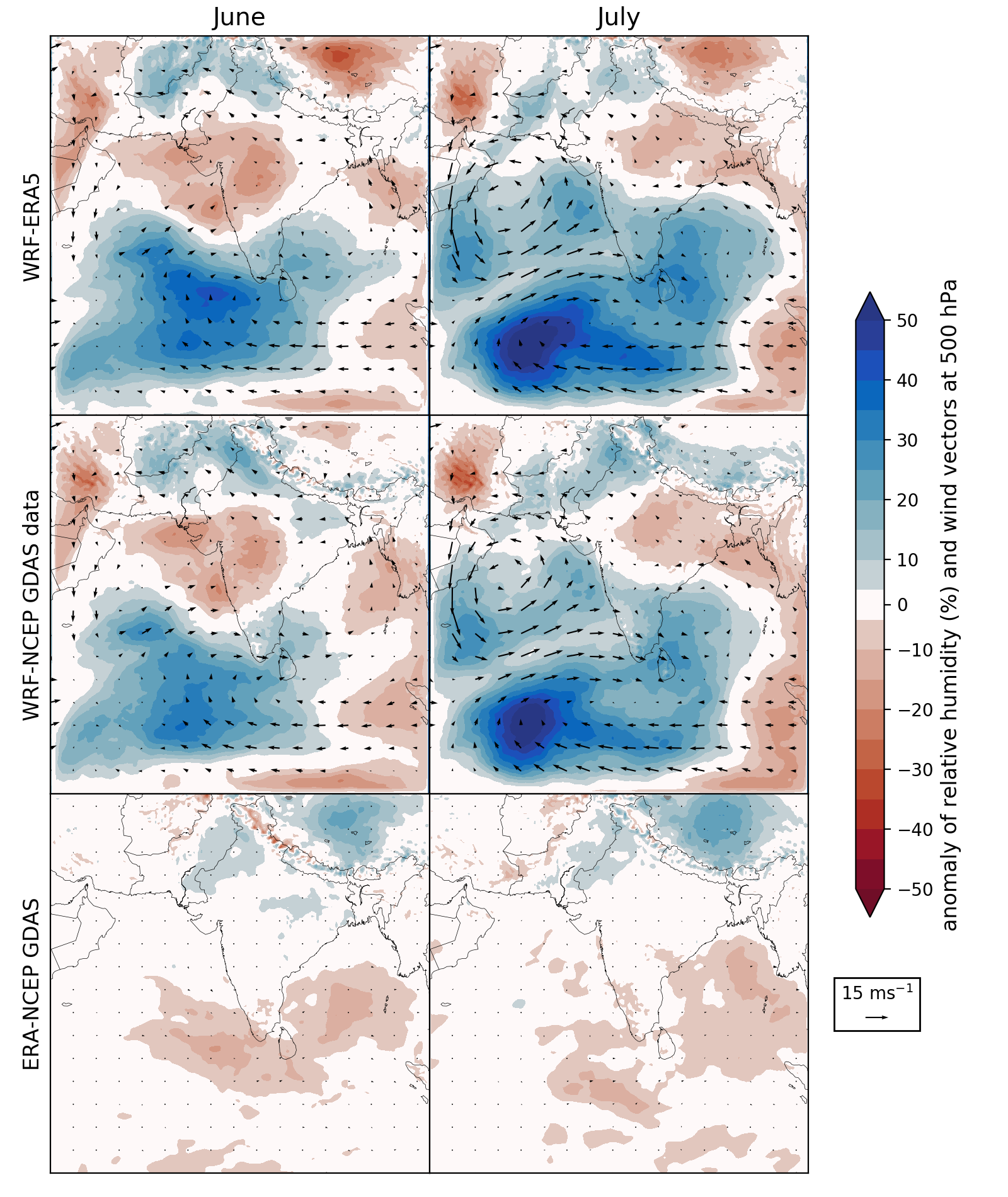} 
\caption{Anomaly of monthly averaged relative humidity (\%) and wind vectors (ms\textsuperscript{-1}) at 500 hPa for June \& July 2016 between the WRF model \& ERA5 reanalysis data (top row), the WRF model \& NCEP GDAS/FNL reanalysis data (middle row) and ERA5 \& NCEP GDAS/FNL reanalysis data (bottom row).}
\label{figureS05}
\end{figure}

\begin{figure}[!ht]
\centering
\includegraphics[width=12cm]{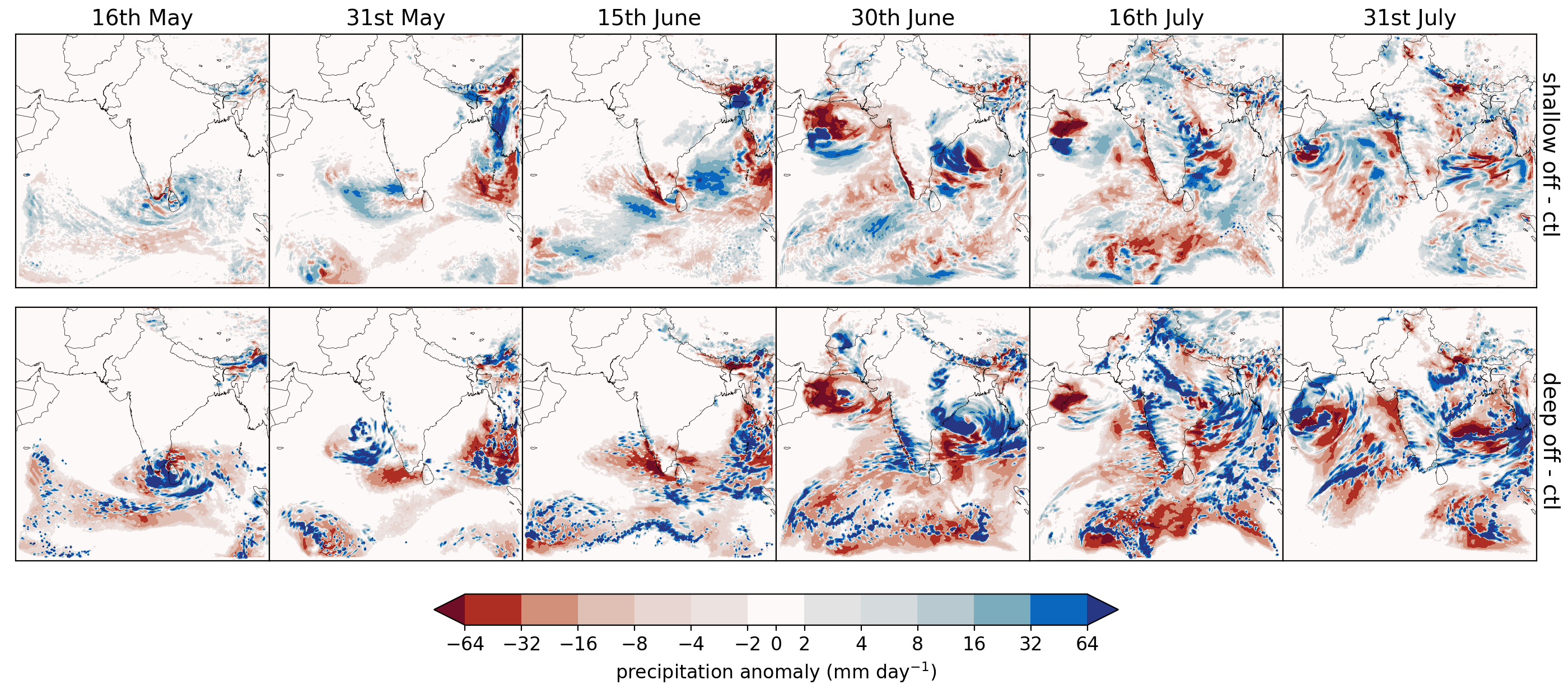} 
\caption{Anomaly of daily accumulated precipitation (mm/day) for various dates from the 11 week simulations with the WRF model. Anomaly of shallow \& mid level convection switched off (top), and deep convection switched off (bottom), compared to the control run.}
\label{figureS06}
\end{figure}

\begin{figure}[!ht]
\centering
\includegraphics[width=12cm]{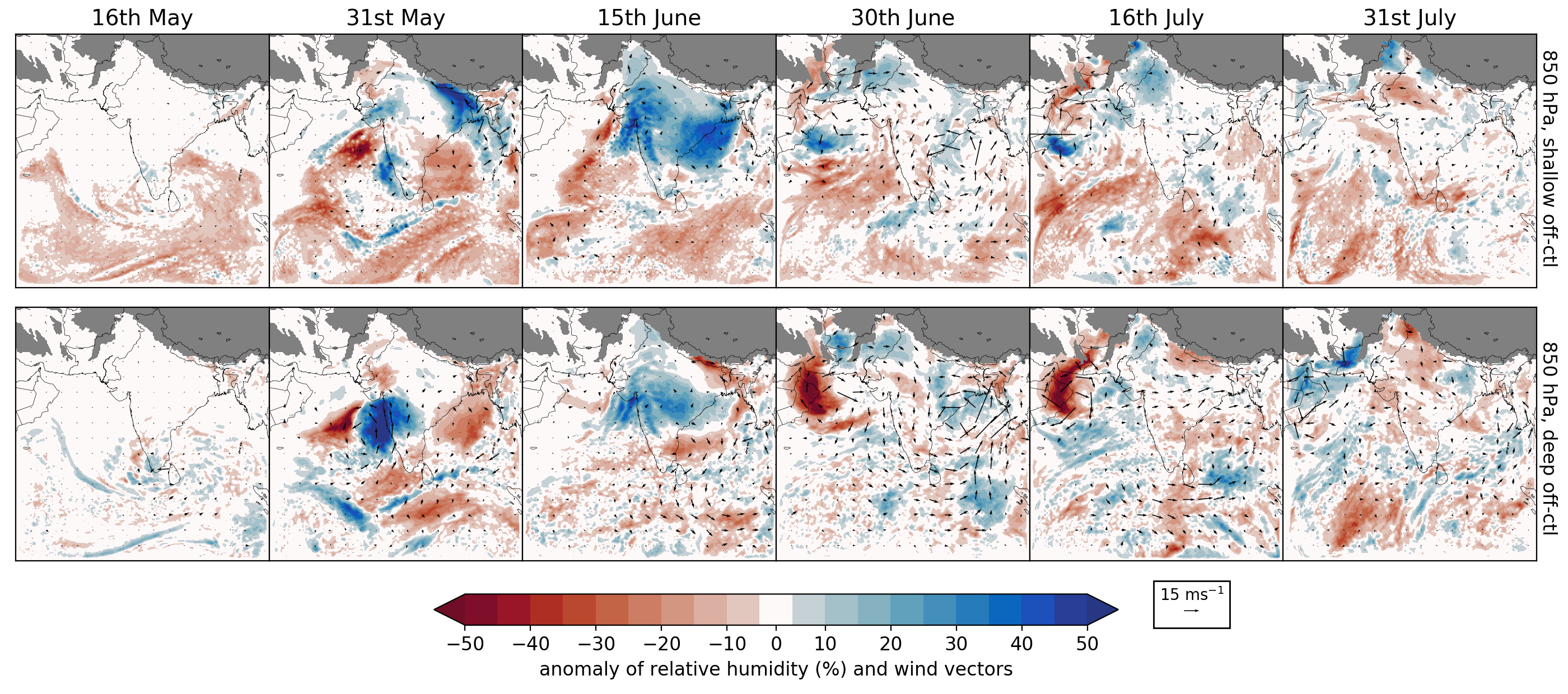}  
\caption{Anomaly of relative humidity (shading, \%) and wind (vectors, ms\textsuperscript{-1}) at 850 hPa, for various dates from the 11 week simulations with the WRF model. Anomaly of shallow \& mid level convection switched off (top), and deep convection switched off (bottom), compared to the control run.}
\label{figureS07}
\end{figure}

\begin{figure}[!ht]
\centering
\includegraphics[width=12cm]{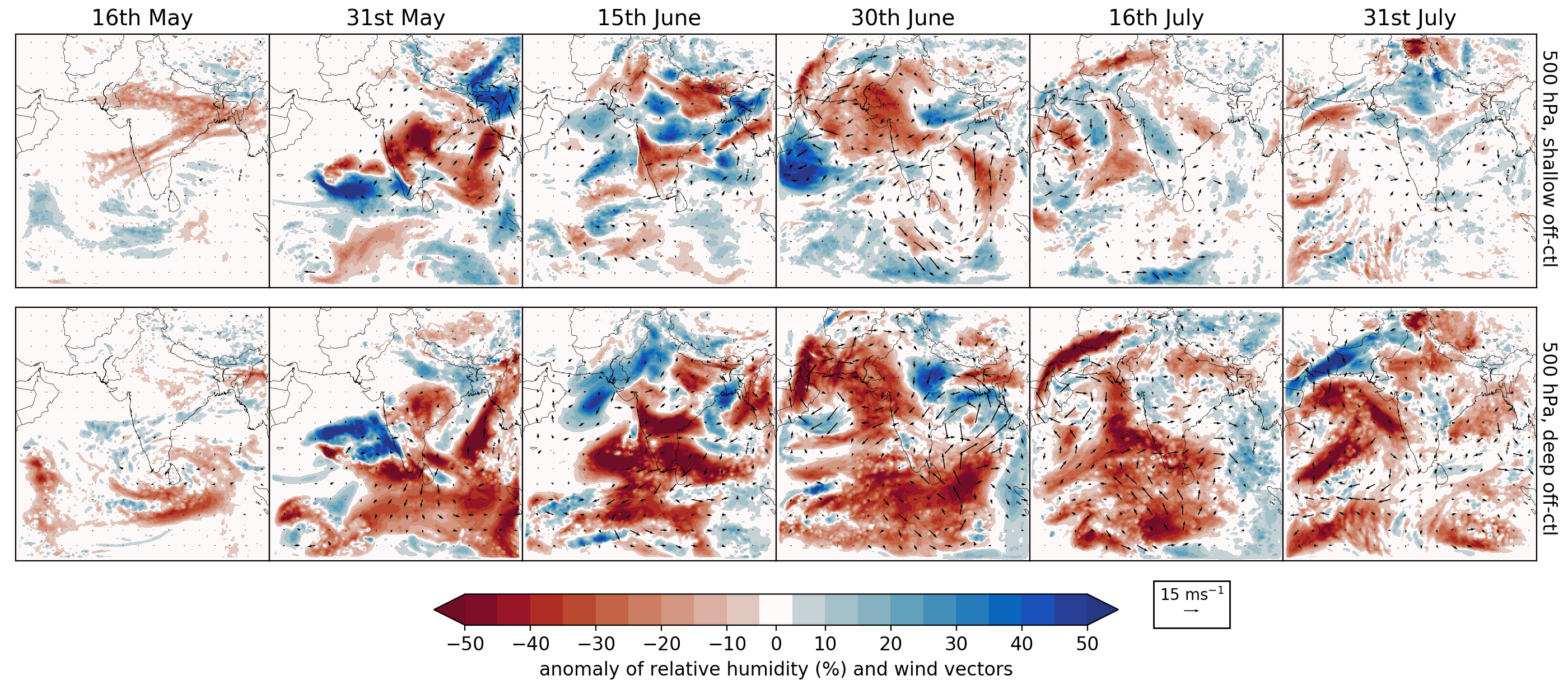} 
\caption{Anomaly of relative humidity (shading, \%) and wind (vectors, ms\textsuperscript{-1}) at 500 hPa, for various dates from the 11 week simulations with the WRF model. Anomaly of shallow \& mid level convection switched off (top), and deep convection switched off (bottom), compared to the control run.}
\label{figureS08}
\end{figure}

\begin{figure}[!ht] 
\centering 
\includegraphics[width=12cm]{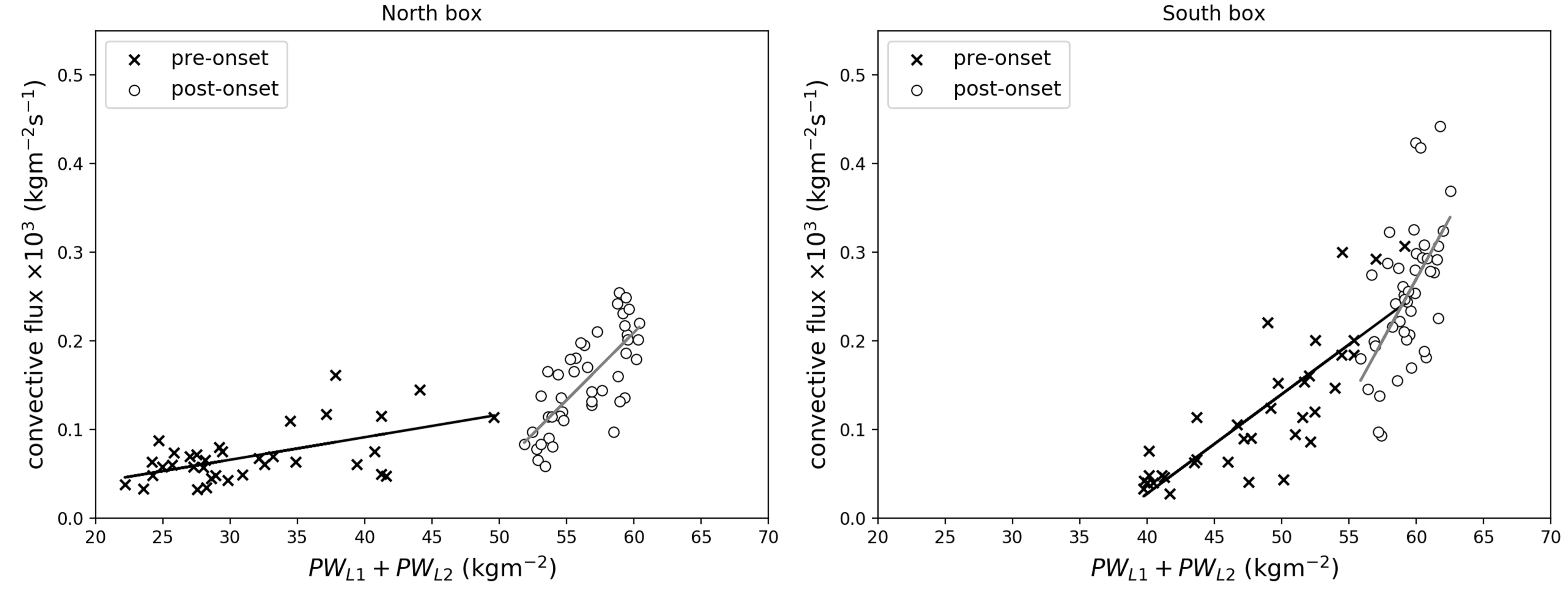} 
\caption{Correlation of vertical convective flux with the sum of the integrated moisture content in the lower and upper layers. Pre-onset refers to the period before 18th June, and post-onset the period after. Daily averaged data from 11 week simulation with the WRF model with shallow \& mid level convection switched off, over the north \& south boxes shown in Figure \ref{figure01}.}
\label{figureS09}
\end{figure}

\begin{figure}[!ht]
\centering 
\includegraphics[width=12cm]{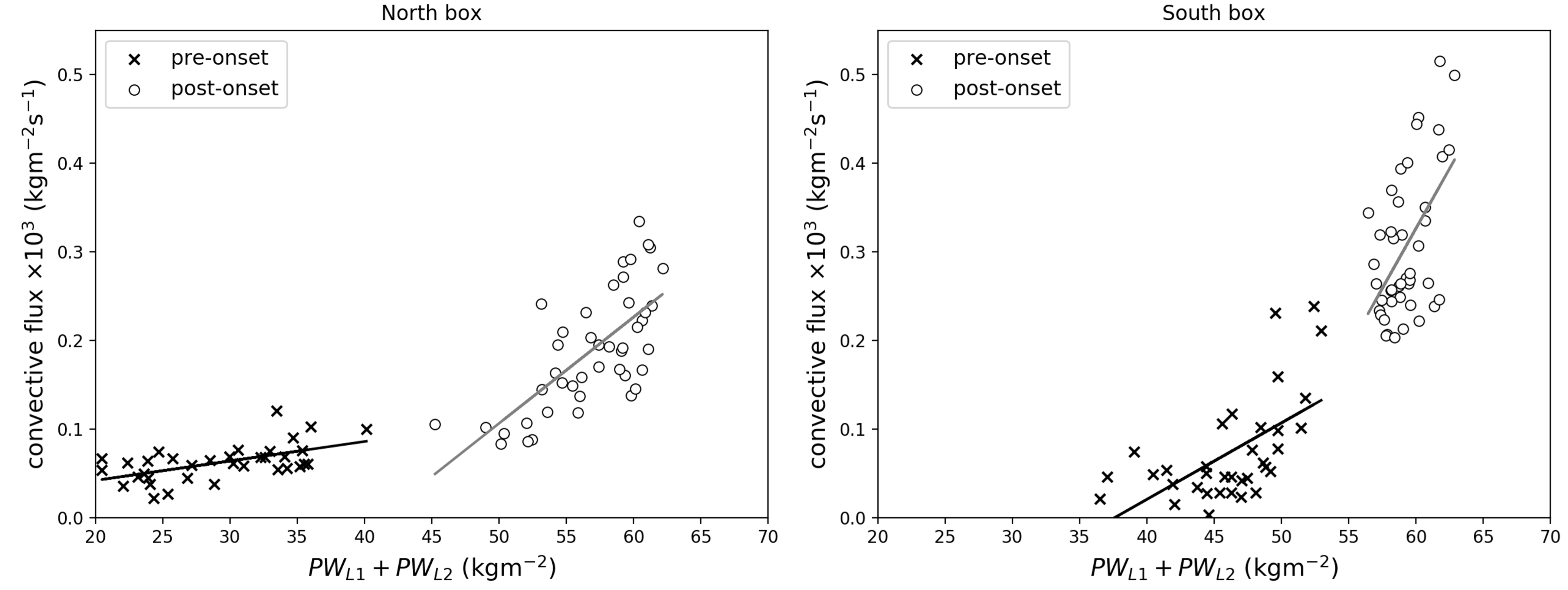}  
\caption{Correlation of vertical convective flux with the sum of the integrated moisture content in the lower and upper layers. Pre-onset refers to the period before 18th June, and post-onset the period after. Daily averaged data from 11 week simulation with the WRF model with deep convection switched off, over the north \& south boxes shown in Figure \ref{figure01}.}
\label{figureS10}
\end{figure}

%-----------------------------------------------------------------------------------------------

\end{document}